\newtheorem{theorem}{Theorem}
\newtheorem{proposition}{Proposition}

\newtheorem{corollary}{Corollary}

\documentclass[12pt]{article}

\usepackage{amssymb}
\usepackage{amsfonts}
\usepackage{mathrsfs}
\usepackage{graphicx}
\usepackage{amsmath}
\usepackage{subfigure}

\usepackage{tabularx}
\usepackage{booktabs}
\usepackage{caption}

\newcommand{\qed}{{\mbox{} \hspace*{\fill}{\vrule height5pt width4pt depth0pt}}\\}

\begin{document}
\bibliographystyle{plain}
\title{A Partial Order for Strictly Positive Coalitional Games and a Link from Risk Aversion to Cooperation}
\author{Jian Yang\\
Department of Management Science and Information Systems\\
Business School, Rutgers University, Newark, NJ 07102}

\date{April 2023}

\maketitle
	
\begin{abstract}
	
We deal with coalitional games possessing strictly positive values. Individually rational allocations of such a game has clear fractional interpretations. Many concepts, including the long-existing core and other stability notions more recently proposed by Yang \cite{Y22}, can all be re-cast in this fractional mode. The latter allows a certain ranking between games, which we deem as in the sense of ``centripetality'', to imply a clearly describable shift in the games' stable solutions. 
When coalitions' values are built on both random outcomes and a common positively homogeneous reward function characterizing players' enjoyments from their shares, the above link could help explain why aversion to risk often promotes cooperation.     
	
\vspace*{3mm} \noindent{\bf Keywords: }Coalitional Game; Stability; Partial Order; Risk Aversion; Cooperation

\vspace*{.1in}\noindent{\bf JEL Code: }C71
\end{abstract}

\newpage

\section{Introduction}\label{introduction}

When faced with the phenomena of lawyers joining firms, artisan shops forming guilds, and nations entering alliances, one cannot help but wondering that perhaps a main driver behind all these cooperations is the instinct to avert risks, one that is probably as primal as the urge to earn more profits on the average.  Our ultimate purpose in this study is to provide some concrete evidence to support this speculation in the formal framework of the coalitional game. 
Such an attempt 
was already made by Yang and Li \cite{YL20}. Due probably to their focus on the traditional concept of core whose nonemptiness comes with strings attached, results there were not as crisp as desired.

Recently, Yang \cite{Y22} proposed stability notions that are both universal in the sense that all games possess their solutions deemed stable by the notions and also compatible with, though not straightforwardly the core, but their weakened varieties. One of the weakening products is the weak core. While members of the traditional core need to withstand attempted changes of the unilateral type that benefit just some or even only one sub-coalition, the weak core would require members to only survive breaking attempts that benefit all parties. This principle of consensual blocking was applied to a weak stability notion $\mathbb S^-$ that involves not only allocation vectors $\vec x$ as was already the case with the core concepts but also partitions ${\cal P}$, those that divide the grand coalition $N$ into constituent coalitions $C$. Certainly, ${\cal P}$ is not necessarily the all-consolidated partition $\{N\}$, the traditional center of attention. After all, the ``natural resting state'' of an arbitrary game might be somehow splintered. 



Here, we leverage these stability notions to advance on the same mission taken up earlier by Yang and Li \cite{YL20}. Technical details being set aside, even the notion of ``more cooperation'' has undergone a generalization. Earlier, it meant the enlargement of the core which is itself not always nonempty; now, it means the march to the ``more consolidated'' end of stable partitions which are always in existence for the weak case and at each such partition, the enlargement of choices on stable allocations. 
During this process, we find it more convenient to view allocations as fractions $f$ rather than absolute values $x$. 

The fractional view would be valid when we are confined to games with strictly positive values $v(C)$, where the strictness can be relaxed when $|C|=1$. 
Though not serious for quite some applications, we still hope future research could help remove the various positivity restrictions. Now for the slightly restricted games, we propose a partial order which we call ``centripetality''. A game ranked higher by this order would have a higher ratio $v(C^2)/v(C^1)$ when $C^1\subseteq C^2$; see Figure~\ref{fig1} later. Our logical flow is
\begin{equation}\label{logic}
\mbox{Risk Aversion}\;\;\Longrightarrow\;\;\;\mbox{Centripetality}\;\;\;\Longrightarrow\;\;\;\mbox{Consolidation\;/\;Cooperation}.
\end{equation}
We shall, however, first work out the second link in~(\ref{logic}). 

The centripetality order would play a clear-cut role when Yang's \cite{Y22} fission- and fusion-resistance notions are recast in the current fractional mode. Here, a solution $({\cal P},\vec f)$ consists of a partition ${\cal P}$ 
and a fractional allocation $\vec f\equiv (f(i))_{i\in N}$ with each $C$-portion $\vec f\,|_C\equiv (f(i))_{i\in C}$ providing a plan to divide the corresponding value $v(C)$. There are strong, medium, and weak fission resistances. For instance, the given pair would be weakly fission-resistant when no fission product ${\cal P}'$ of ${\cal P}$  could make all involved players strictly happier. 
Fusion resistance can be symmetrically defined. That ``weak'' or any of the three adjectives need not be applied to the latter stems from a fundamental asymmetry between the symmetrically-sounding fission and fusion; we shall explain this later. Now, a member $\vec x^{\,*}$ of the weak core, when fractionally reinterpreted as some $\vec f^{\,*}$, could be understood as one that makes $(\{N\},\vec f^{\,*})$ weakly fission-resistant. Certainly, any $(\{N\},\vec f)$ is already fusion-resistant. 

Our main result Theorem~\ref{t-relation} states that, when a game ascends in the centripetality order, it would make a given solution $({\cal P},\vec f)$ ``easier'' to be fission-resistant of the weak and strong categories and ``harder'' to be fusion-resistant. As a given game is already ``easier'' for a more splintered partition to be fission- rather than fusion-resistant and for a more consolidated partition to be fusion- rather than fission-resistant, the overall effect of more centripetality is that stable solutions become ``more consolidated''; consult Figure~\ref{fig2} later. When it comes to the already all-consolidated partition $\{N\}$, increases in centripetality would make it ``more likely'' for the core of any category to emerge and when nonempty, for it to grow ``larger''; see Corollary~\ref{c-extremes} which addresses this and the symmetric, most splintered, case.  

We shall omit Yang's \cite{Y22} medium category as, unlike the weak and strong ones, it does not seem to induce the above monotone comparative static results. After thus completing the second link in~(\ref{logic}), we move on to the first link.  
In this part, suppose each coalition $C$ is associated with a random outcome $\Phi(C)$ and all players use one common positively homogeneous reward function $\tilde R$ to assess their enjoyments over the shares allotted to them. Then, the link can be established for two scenarios. 

In the first scenario, each $\Phi(C)$ is deterministically proportional to the sum of the $C$-participants' individual inputs which are independent and identically distributed; also, $\tilde R$ is the expectation less some $\bar r$ times the standard deviation. Players would become more risk averse when $\bar r$ increases. In the second scenario, the random outcomes $\Phi(C)$ are expressed through their quantile functions that satisfy certain conditions reflecting larger coalitions' relatively smaller tails; $\tilde R$ reflects a mixture of conditional values at risk. Players would be more risk averse when the mixing density $\bar\mu$ on the risk level increases in the likelihood-ratio sense. For both scenarios, we demonstrate that ``aversion to risk boosts centripetality''; see Propositions~\ref{p-pity} and~\ref{p-okla}. Then, as indicated by~(\ref{logic}), Theorem~\ref{t-relation} and Corollary~\ref{c-extremes} would be conduits through which more risk aversion and more cooperation are linked up. 

Besides Yang and Li \cite{YL20}, there does not seem to be many other attempts on attributing coalition formations to risk attitudes. Some works did introduce randomness to cooperative games; see, e.g., Charnes and Granot \cite{CG73}\cite{CG76}\cite{CG77}, Suijs et al. \cite{Suijs et al. 1999}, and Suijs and Borm \cite{SB99}. To the best of our knowledge, however, these works did not touch on the changes of players' risk attitudes. Also related might be the literature started by Wilson \cite{W78} on information asymmetry; see, e.g., Allen and Yannelis \cite{AY01}, Forges, Minelli, and Vohra \cite{FMV02}, and Myerson \cite{M07}. Our current model is not yet concerned with the phenomenon of different players possessing different signals of an underlying value-affecting state. 

We hope future research will bring further improvements to both links in~(\ref{logic}). On the one hand, it would be ideal to have the strict-positivity requirement on values removed; on the other hand, more general and broader cases of risk aversion might still be involved. As it currently stands, this paper is organized as follows. We establish the second link in four sections, with Section~\ref{fractional} introducing the fractional form and Section~\ref{s-stability} re-casting Yang's \cite{Y22} stability notions in the current fractional framework, Section~\ref{ordering} detailing the centripetality order and deriving its main consequences on consolidation and cooperation, and Section~\ref{implications} working on further implications. The first link in~(\ref{logic}) is formed for two cases, which we work out in Sections~\ref{inoutreward} and~\ref{situ2}, respectively. The paper is concluded in Section~\ref{conclusion}. 

\section{The Fractional Form}\label{fractional}

We consider the traditional transferable utility (TU) game $(N,\vec v)$ where $N$ is the finite set of players which spurs the set $\mathscr C(N)$ of its nonempty subsets or coalitions; also, $\vec v\equiv (v(C))_{C\in \mathscr C(N)}$ represents the vector of coalitions' values. The only restriction we shall place here is that each $v(\{i\})$ be in $\Re_+$, the set of positive reals and each $v(C)$ for $|C|\geq 2$ be in $\Re_{++}$, the set of strictly positive reals. For convenience, use $\mathscr V_{++}(N)$ for the set of all strictly positive games; basically, it is $\prod_{i\in N}\Re_{+}\times\prod_{C\in\mathscr C(N),\;|C|\geq 2}\Re_{++}$. 

As it is always possible to transform an ordinary game into a strictly positive one without altering important properties like the nonemptiness of the core, most authors would not consider this restriction to be serious. 
Let us then focus on a $\vec v\in \mathscr V_{++}(N)$. Ordinarily, an individually rational and efficient allocation $\vec x\equiv (x(i))_{i\in N}$ should satisfy
\[ x(i)\geq v(\{i\})\mbox{ for every }i\in N\;\;\;\mbox{ and }\;\;\;\sum_{i\in N}x(i)=v(N).\]
As $v(\{i\})\geq 0$ and $v(N)>0$ when $|N|\geq 2$, an alternative understanding of the allocation is some positive vector $\vec f\equiv (f(i))_{i\in N}$ with $\sum_{i\in N}f(i)=1$; indeed, $f(i)=1$ when $N=\{i\}$ and $f(i)=x(i)/v(N)$ for every $i\in N$ when $|N|\geq 2$ would facilitate the transition. 

Thus, define the set $\partial(N,\vec v)$ of individually rational and efficient fractional allocations by  
\begin{equation}\label{simplex-def}
\left\{\begin{array}{ll}
\{1\} & \;\;\;\;\;\;\mbox{ when }|N|=1,\\
\left\{\vec f\equiv (f(i))_{i\in N}\in \prod_{i\in N} [v(\{i\})/v(N),1]:\;\sum_{i\in N}f(i)=1\right\} & \;\;\;\;\;\;\mbox{ when }|N|\geq 2.
\end{array}\right.\end{equation}
Like Yang \cite{Y22}, we focus on not only the grand coalition but also more generally, the partition-based stable solutions. Let $\mathscr P(N)$ be the set of partitions that can be formed out of players in $N$. Each partition ${\cal P}\in \mathscr P(N)$ would be of the form $\{C_1,...,C_p\}$, where the $C_k$'s are non-overlapping members of $\mathscr C(N)$ that together make up the grand coalition $N$. Let 
\begin{equation}\label{allocation-def}
\mathscr F(N,\vec v,{\cal P})\equiv \prod_{C\in {\cal P}}\partial(C,\vec v\,|_{\mathscr C(C)}),
\end{equation}
which is the partition's set of fractional allocation plans that are feasible in the sense that they are individually rational and for each constituent coalition $C\in {\cal P}$, the sub-plan $\vec f\,|_{C}\equiv (f(i))_{i\in C}$ is efficient for $C$. In~(\ref{allocation-def}), each $(C,\vec v\,|_{\mathscr C(C)})$ certainly refers to the strictly positive sub-game composed of the player set $C$ and payoff vector $\vec v\,|_{\mathscr C(C)}\equiv (v(C'))_{C'\in\mathscr C(C)}$. 

Note $\mathscr F(N,\vec v,\{N\})=\partial(N,\vec v)$. For at least the all-splintered partition $\underline{\cal P}\equiv \{\{i\}:\;i\in N\}$, the feasible set $\mathscr F(N,\vec v,\underline{\cal P})$ of~(\ref{allocation-def}) is nonempty as it includes the all-one vector $\vec 1$. We consider the following to be the game $(N,\vec v)$'s feasible set of solutions:
\begin{equation}\label{pair-def}
\mathscr Q(N,\vec v)\equiv \bigcup_{{\cal P}\in \mathscr P(N)}\{{\cal P}\}\times \mathscr F(N,\vec v,{\cal P}).
\end{equation}
By~(\ref{simplex-def}) to~(\ref{pair-def}), a partition-allocation pair $({\cal P},\vec f)$ would be considered feasible when $\vec f$'s $C$-portion $\vec f\,|_C\equiv (f(i))_{i\in C}$, for every constituent coalition $C$ of the partition ${\cal P}$, constitutes an individually rational plan to apportion the coalition's value. Note $(\underline{\cal P},\vec 1)\in\mathscr Q(N,\vec v)$ and hence the latter is always nonempty. 

The traditional core in the fractional mode, which could be understood as those vectors $\vec f$ that render partition-allocation pairs $(\{N\},\vec f)$ stable in some sense, is clearly
\begin{equation}\label{core-def-trad0}
\mathbb F^+(N,\vec v)\equiv\left\{\vec f\in \partial(N,\vec v):\;v(N)\cdot\sum_{i\in C}f(i)\geq v(C),\;\;\;\forall C\in \mathscr C(N)\hspace*{-.02in}\setminus\hspace*{-.03in}\{N\}\right\}.
\end{equation}
For convenience, we still impose in~(\ref{core-def-trad0}) inequalities that are redundant to those already imposed by~(\ref{simplex-def})'s individual-rationality requirements. 


\section{Stability for Partition-allocation Pairs}\label{s-stability}

The spirit reflected in~(\ref{core-def-trad0}) could still be questioned. Consider an example from Yang \cite{Y22}. Suppose Alice is in an unhappy marriage with Bob with their daughter Carol caught in the middle. According to the old spirit, Alice could well break away from the current union and take Carol along with her as long as she and her daughter is strictly happier afterwards; this is without any regard for Bob's welfare. Reality might be more complex. Very likely, Bob would try hard to block Alice and Carol from leaving him unless his post-marriage life is well taken care of as well. This motivates us to weaken~(\ref{core-def-trad0}).

Note it equates $\mathbb F^+(N,\vec v)$ to
\begin{equation}\label{core-def-trad}
\left\{\vec f\in \partial(N,\vec v):\;\forall C\in {\cal P}\mbox{ we have }v(N)\cdot \sum_{i\in C}f(i)\geq v(C),\;\;\forall {\cal P}\in \mathscr P(N)\hspace*{-.02in}\setminus\hspace*{-.03in}\{\{N\}\}\right\}.
\end{equation}
By~(\ref{core-def-trad}), it takes $\vec f\,|_C\equiv (f(i))_{i\in C}$'s resistance against $v(C)$ at all coalitions $C$ or equivalently, all constituent coalitions of any non-$\{N\}$ partition, for the fractional vector $\vec f$ to be stable in the traditional sense. Based on the Alice-Bob-Carol story, however, it might be reasonable to weaken~(\ref{core-def-trad}) into the following weak core in a fractional form or weak fractional core $\mathbb F^-(N,\vec v)$:
\begin{equation}\label{core-def-weak}
\left\{\vec f\in \partial(N,\vec v):\;\exists C\in {\cal P}\mbox{ so that }v(N)\cdot\sum_{i\in C}f(i)\geq v(C),\;\;\forall {\cal P}\in \mathscr P(N)\hspace*{-.02in}\setminus\hspace*{-.03in}\{\{N\}\}\right\}.
\end{equation}
By~(\ref{core-def-weak}), it would take the same resistance at merely one representative coalition of each non-$\{N\}$ partition, for $\vec f$ to be stable in our weak sense. Think of the stabilizing role played by $\{\mbox{Bob}\}$ in the unhappy but potentially real-life-sustainable marriage.  

In the sense that both condone self-centered sub-coalitions to hold everyone else hostage,~(\ref{core-def-trad}) and~(\ref{core-def-weak}) are rather alike. They are only different on the guiding principle: whether it should be a prized state that is ``hard to earn when not yet acquired'' or rather a stable one that is ``difficult to part with once already in possession''. By~(\ref{core-def-trad}), any single existing sub-coalition of players fretting about their futures together with others could sabotage the fusion into the grand coalition; by~(\ref{core-def-weak}), any single sub-coalition of players concerned of their fates when all alone by themselves could stop the fission of the already-formed grand coalition. 

According to Yang \cite{Y22}, there is also a plausible middle ground, the medium core $\mathbb F^0(N,\vec v)$. Since it does not seem to induce the same monotone comparative statics as those achieved by its weak and strong counterparts, we omit presenting it here.  
Now, as might be expected from~(\ref{core-def-trad}) and~(\ref{core-def-weak}), it is indeed true that
\begin{equation}\label{core-relations}
\mathbb F^+(N,\vec v)\subsetneq
\mathbb F^-(N,\vec v).
\end{equation}

Now let us move beyond the grand-coalition-obsessed $(\{N\},\vec f)$'s where the $\vec f$'s belong to some core to general stable solutions $({\cal P},\vec f)$, where the ${\cal P}$'s are by no means just $\{N\}$. Following the spirits of~(\ref{core-def-trad}) and~(\ref{core-def-weak}), we shall consider a solution $({\cal P},\vec f)$ belonging to $\mathscr Q(N,\vec v)$ of~(\ref{pair-def}) as stable to a certain degree for the game $(N,\vec v)$ when it attains both fission resistance to a certain degree and also fusion resistance. 

First, let $\mathscr I(N,{\cal P})$ be ${\cal P}$'s fission-down-to neighborhood whose every member ${\cal P}'$ is formed by sub-coalitions of ${\cal P}$'s constituent coalitions. Also, let $\mathscr U(N,{\cal P})$ be ${\cal P}$'s fusion-up-to neighborhood whose every member ${\cal P}'$ is formed by unions of ${\cal P}$'s constituent coalitions. For instance, $\mathscr I(\{1,2,3\},\{\{1\},\{2,3\}\})$ contains only one member, the all-splintered partition $\underline{\cal P}\equiv\{\{1\},\{2\},\{3\}\}$; meanwhile, $\mathscr U(\{1,2,3\},\{\{1\},\{2,3\}\})$ also contains only one member, the all-consolidated partition $\{N\}\equiv \{\{1,2,3\}\}$. 

Due to a subtlety to be emphasized, let us present weak fission resistance for a solution $({\cal P},\vec f)\in \mathscr Q(N,\vec v)$ first. For the game $(N,\vec v)$, it means for any ${\cal P}'\in \mathscr I(N,{\cal P})$, 
\begin{equation}\label{stability-def}
\exists C'\in {\cal P}'\hspace*{-.02in}\setminus\hspace*{-.03in}{\cal P}\;\;\;\mbox{ so that }\;\;\;\sum_{C\in {\cal P}\mbox{ and }C\supsetneq C'}v(C)\cdot \sum_{i\in C'}f(i)\geq v(C').
\end{equation}
By~(\ref{stability-def}), any attempt to further split a weakly fission-resistant pair $({\cal P},\vec f)$ should receive the objection from at least one affected sub-coalition $C'$. The requirement that $C'\notin {\cal P}$ ensures that the $C'$ in existence is a sub-coalition in the strict sense. When ${\cal P}$ happens to be $\{N\}$, a constituent coalition $C'$ of any partition ${\cal P}'$ in the fission-down-to neighborhood $\mathscr I(N,\{N\})$ would not be $N$ itself. Then, the requirement $C'\in {\cal P}'\hspace*{-.02in}\setminus\hspace*{-.03in}\{N\}$ would be the same as $C'\in {\cal P}'$. This may help us see that~(\ref{core-def-weak}) is consistent with~(\ref{stability-def}).  

Note there is only one coalition $C$ satisfying both $C\in {\cal P}$ and $C\supsetneq C'$: the unique coalition that strictly contains $C'\in {\cal P}'\hspace*{-.02in}\setminus\hspace*{-.03in}{\cal P}$. Hence,~(\ref{stability-def}) could well be expressed as 
\begin{equation}\label{fi-resist-weak}
\exists C'\subsetneq C\;\mbox{ so that }\;v(C)\cdot\sum_{i\in C'}f(i)\geq v(C'),\hspace*{.8in}\forall C\in {\cal P}\mbox{ with }|C|\geq 2.
\end{equation}
According to~(\ref{stability-def}) in the current definition or namely just~(\ref{fi-resist-weak}), for at least one strict sub-coalition $C'$ of any constituent $C$ of the given partition ${\cal P}$, staying put in $C$ is beneficial.   

In the spirit of~(\ref{core-def-trad}), we deem a $({\cal P},\vec f)\in\mathscr Q(N,\vec v)$ strongly fission-resistant when 
\begin{equation}\label{fi-resist-strong}
\forall C'\subsetneq C\;\mbox{ we have }\;v(C)\cdot\sum_{i\in C'}f(i)\geq v(C'),\hspace*{.8in}\forall C\in {\cal P}\mbox{ with }|C|\geq 2.
\end{equation}

A split affects not only players in a targeted sub-coalition but also those whose participation might be deemed unwilling; meanwhile, a merger involves only players in the targeted super-coalition. We might understand the traditional core as tolerant of collateral damages caused by splittings and the weak core as intolerant. But for mergers, there is no collateral damage to speak of. Therefore, let us consider just one plain fusion resistance. 

For any $C'$ that is a strict merger of ${\cal P}$'s constituent coalitions, there could be multiple $C$'s that satisfy $C\in {\cal P}$ and $C\subsetneq C'$: it is their union that helps to form $C'$. Hence, each $\sum_{i\in C}f(i)=1$ due to the feasibility of $({\cal P},\vec f)$ and the definitions~(\ref{simplex-def}) to~(\ref{pair-def}). We can thus express fusion resistance as for any ${\cal P}'\in \mathscr U(N,{\cal P})$, 
\begin{equation}\label{fu-resist}
\forall C'\in {\cal P}'\hspace*{-.02in}\setminus\hspace*{-.03in}{\cal P}\;\;\;\mbox{ we have }\;\;\;\sum_{C\in {\cal P}\mbox{ and }C\subsetneq C'}v(C)\geq v(C').
\end{equation}
The current~(\ref{fu-resist}) means that by forming any strict  super-coalition $C'$ of some constituent coalitions $C$ of the given partition ${\cal P}$, no player in $C'$ could hope to be strictly better off without hurting any other player in the super-coalition. 
Unlike the two fission resistances, fusion resistance is a solely-${\cal P}$-determined property. 

\section{The Centripetality Partial Order}\label{ordering}

Because there is no substantial difference between the current fractional form and the traditional one, we can inherit many notions from Yang \cite{Y22}. However, it is ultimately the new form that would help us achieve a meaningful partial ordering among strictly positive games. 

For any $v\in\mathscr V_{++}(N)$, define $\mathbb Q^{\mbox{i}\pm}(N,\vec v)\subset \mathscr Q(N,\vec v)$ of~(\ref{pair-def}) as the set of solutions $({\cal P},\vec f)$ that are fission-resistant in the sense of~(\ref{fi-resist-weak}) or~(\ref{fi-resist-strong}). By these definitions, a solution $({\cal P},\vec f)$'s stability is equated to its membership to 
\begin{equation}\label{stab-def}
\mathbb S^\pm(N,\vec v)\equiv \mathbb Q^{\mbox{i}\pm}(N,\vec v)\cap \mathbb Q^{\mbox u}(N,\vec v).
\end{equation} 
Furthermore, let $\mathbb Q^{\mbox{u}}(N,\vec v)\subset \mathscr Q(N,\vec v)$ be the set of solutions $({\cal P},\vec f)$ that are fusion-resistant in the sense of~(\ref{fu-resist}). 
The latter's $\vec f$-independence implies the existence of some $\mathbb P^{\mbox{u}}(N,\vec v)$ for
\begin{equation}\label{qu}
\mathbb Q^{\mbox{u}}(N,\vec v)=\bigcup_{{\cal P}\in \mathbb P^{\mbox{u}}(N,\vec v)}\{{\cal P}\}\times\mathscr F(N,\vec v,{\cal P}).
\end{equation}

\begin{figure}[h]
\centering
\includegraphics[width=0.85\columnwidth]{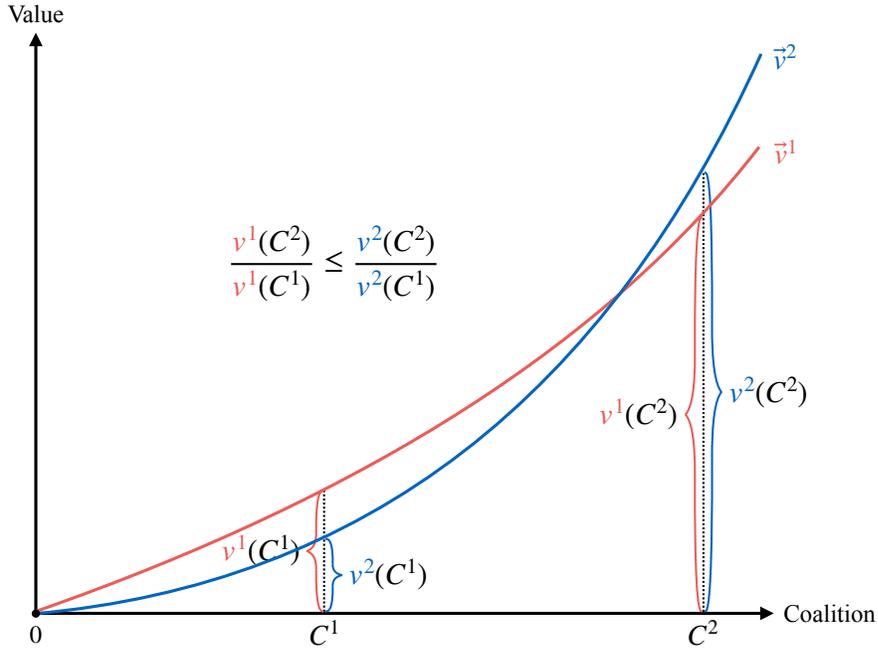}
\caption{An Illustration of the Centripetality Order}\label{fig1}
\end{figure}

The definitions around~(\ref{simplex-def}) to~(\ref{pair-def}) and those around~(\ref{fi-resist-weak}) to~(\ref{fu-resist}) prompt us to introduce a partial order among strictly positive games. Given $\vec v^{\,1},\vec v^{\,2}\in \mathscr V_{++}(N)$, we deem $\vec v^{\,1}$ to be less centripetal than $\tilde v^{\,2}$, denoted as either $\vec v^{\,1}\leq_{\mbox{cp}} \vec v^{\,2}$ or $\vec v^{\,2}\geq_{\mbox{cp}} \vec v^{\,1}$, when 
\begin{equation}\label{partial-order}
\frac{v^1(C^2)}{v^1(C^1)}\leq \frac{v^2(C^2)}{v^2(C^1)},\hspace*{.8in}\forall C^1,C^2\in \mathscr C(N)\mbox{ with }C^1\subseteq C^2. 
\end{equation}
We might let~(\ref{partial-order}) adopt the convention that $0/0=1$, $a/0=+\infty$ for $a>0$, and $+\infty=+\infty$ for it to accommodate the possibility that $|C^i|=1$ and $v^j(C^i)=0$ for $i,j=1,2$. Figure~\ref{fig1} gives an illustration to this centripetality order. 

From Figure~\ref{fig1}, we may see that $\vec v^{\,1}\leq_{\mbox{cp}}\vec v^{\,2}$ is a reflection of $\vec v^{\,1}$'s enhanced appreciation of smaller sets and $\tilde v^{\,2}$'s enhanced appreciation of larger sets. 
If so, we should expect a solution pair $({\cal P},\vec f)$ that is fission-resistant under $\vec v^{\,1}$ to remain so under $\tilde v^{\,2}$ and symmetrically, a solution pair $({\cal P},\vec f)$ that is fusion-resistant under $\vec v^{\,2}$ to remain so under $\vec v^{\,1}$. 

A namesake of centripetality was defined by Yang and Li \cite{YL20} as well. In the core-centric study there, it referred to the weaker version where $C^2=N$. Its implications were much more limited as well. Besides the confinement to the traditional core, it did not lead to such a clear-cut second link of~(\ref{logic}) as to be done here either. From Theorem 3 of Yang and Li \cite{YL20}, one might see that centripetality there and a special version of risk aversion were very much entangled. In contrast, our message from the current version of centripetality would be loud and clear thanks in part to recent advances made by Yang \cite{Y22}.  

\begin{theorem}\label{t-relation}
Suppose $\vec v^{\,1},\vec v^{\,2}\in\mathscr V_{++}(N)$ satisfy $\vec v^{\,1}\leq_{\mbox{cp}} \vec v^{\,2}$ in the sense of~(\ref{partial-order}). Then, 
\[ \partial(N,\vec v^{\,1})\subseteq \partial(N,\vec v^{\,2});\]
\[ \mathscr F(N,\vec v^{\,1},{\cal P})\subseteq \mathscr F(N,\vec v^{\,2},{\cal P}),\hspace*{.8in}\forall {\cal P}\in\mathscr P(N);\]
\[ \mathscr Q(N,\vec v^{\,1})\subseteq \mathscr Q(N,\vec v^{\,2});\]
\[ \mathbb Q^{\mbox{i}\pm}(N,\vec v^{\,1})\subseteq \mathbb Q^{\mbox{i}\pm}(N,\vec v^{\,2});\]
\[ \mathbb P^{\mbox u}(N,\vec v^{\,1})\supseteq \mathbb P^{\mbox u}(N,\vec v^{\,2}).\]
\end{theorem} 	

\noindent{\bf Proof of Theorem~\ref{t-relation}: }For the first claim, suppose $\vec f\in \partial(N,\vec v^{\,1})$. By~(\ref{simplex-def}), we have $\vec f=1\in \partial(N,\vec v^{\,2})$ when $|N|=1$. When $|N|\geq 2$, we must have $\sum_{i\in N}f(i)=1$ and each $f(i)\geq v^1(\{i\})/v^1(N)$. But the latter according to~(\ref{partial-order}) is above $v^2(\{i\})/v^2(N)$. Hence, each $f(i)\geq v^2(\{i\})/v^2(N)$ as well. Therefore, $\vec f\in\partial(N,\vec v^{\,2})$. This along with~(\ref{allocation-def}) would result in the second claim, which along with~(\ref{pair-def}) would lead to the third claim. 

For the fourth claim, we first suppose $({\cal P},\vec f)\in \mathbb Q^{\mbox{i-}}(N,\vec v^{\,1})$. According to~(\ref{fi-resist-weak}), this means that for any $C^2\in {\cal P}$ with $|C^2|\geq 2$, there would exist $C^1\subsetneq C^2$ so that 
\begin{equation}\label{kaca}
v^1(C^1)\leq v^1(C^2)\cdot\sum_{i\in C^1}f(i).
\end{equation}
Since~(\ref{partial-order}) is true,~(\ref{kaca}) would imply
\begin{equation}\label{caka}
v^2(C^1)\leq v^1(C^1)\cdot \frac{v^2(C^2)}{v^1(C^2)}\leq v^1(C^2)\cdot\sum_{i\in C^1}f(i)\cdot \frac{v^2(C^2)}{v^1(C^2)}=v^2(C^2)\cdot\sum_{i\in C^1}f(i).  
\end{equation} 
But the existence of a $C^1$ as specified to sustain~(\ref{caka}) for any $C^2$ as specified exactly means that $({\cal P},\vec f)\in \mathbb Q^{\mbox{i-}}(N,\vec v^{\,2})$. Note $\sum_{i\in C^1}f(i)>0$ is derivable but unnecessary here. We have thus completed the derivations for the weak-fission case. In view of the semblance between~(\ref{fi-resist-weak}) and~(\ref{fi-resist-strong}), the strong-fission case can be similarly tackled.

For the last claim concerning the inclusion relationship of $\mathbb P^{\mbox{u}}(N,\cdot)$, suppose ${\cal P}\in \mathbb P^{\mbox{u}}(N,\vec v^{\,2})$ that facilitates~(\ref{qu}). According to~(\ref{fu-resist}), this means that 
\begin{equation}\label{kala}
v^2(C^2)\leq \sum_{C^1\in {\cal P}\mbox{ and }C^1\cap C^2\neq \emptyset}v^2(C^1),
\end{equation}
for any ${\cal P}'\in \mathscr U(N,{\cal P})$ and $C^2\in {\cal P}'\hspace*{-.02in}\setminus\hspace*{-.03in}{\cal P}$. Applying~(\ref{partial-order}) onto~(\ref{kala}) would result in
\begin{equation}\label{laka}\begin{array}{l}
v^1(C^2)=v^2(C^2)\cdot [v^1(C^2)/v^2(C^2)]\\
\;\;\;\;\;\;\;\;\;\leq \sum_{C^1\in {\cal P}\mbox{ and }C^1\cap C^2\neq\emptyset}v^2(C^1)\cdot [v^1(C^2)/v^2(C^2)]\leq  \sum_{C^1\in {\cal P}\mbox{ and }C^1\cap C^2\neq\emptyset}v^1(C^1).  
\end{array}\end{equation} 
But~(\ref{laka}) for any ${\cal P}'\in \mathscr U(N,{\cal P})$ and $C^2\in {\cal P}'\hspace*{-.02in}\setminus\hspace*{-.03in}{\cal P}$ exactly means that ${\cal P}\in \mathbb P^{\mbox u}(N,\vec v^{\,1})$. \qed

In Yang \cite{Y22}, there was a medium stability concept $\mathbb S^0$ built on the medium-core concept $\mathbb F^0$. The universality of the weak stability $\mathbb S^-$ was achieved through that of $\mathbb S^0$. However, this medium concept does not seem to enjoy the monotone comparative static results suggested in  Theorem~\ref{t-relation} simply because $a_1/b_1\leq a_2/b_2$ and $c_1/d_1\leq c_2/d_2$ do not necessarily lead to $(a_1+c_1)/(b_1+d_1)\leq  (a_2+c_2)/(b_2+d_2)$; for instance, $1/1<8/7$ and $2/1<5/2$; yet, $(1+2)/(1+1)=3/2>13/9=(8+5)/(7+2)$.
 
Recall that the absolute-to-fractional transition in the understanding of allocations could be deemed nonessential insofar as  the stability of one particular game is concerned. When it comes to the comparison between different games, however, the transition could be discerned from Theorem~\ref{t-relation}'s proof to be quite indispensable in facilitating a clear-cut monotone result. 

Much more can be said about the theorem itself.

\section{Further Implications of the Main Result}\label{implications}

To understand further implications of Theorem~\ref{t-relation}, let us outline Yang's \cite{Y22} main results regarding the stabilities revolving around~(\ref{fi-resist-weak}) to~(\ref{fu-resist}). 

For a partition ${\cal P}\in \mathscr P(N)$, let the patched-up fractional core $\mathbb F^{\mbox{i}\pm}(N,\vec v,{\cal P})$ be
\begin{equation}\label{patch-up}
\prod_{C\in {\cal P}}\mathbb F^\pm(C,\vec v\,|_{\mathscr C(C)})\equiv\left\{\vec f\in \mathscr F(N,\vec v,{\cal P}):\;\vec f\,|_{C}\in \mathbb F^\pm(C,\vec v\,|_{\mathscr C(C)}),\;\;\forall C\in {\cal P}\right\},
\end{equation}
where each $\mathbb F^\pm(C,\vec v|_{\mathscr C(N)})$ is the fractional core for the corresponding sub-game $(C,\vec v|_{\mathscr C(N)})$ defined at~(\ref{core-def-trad}) or~(\ref{core-def-weak}). Then, define
\begin{equation}\label{pistar-def}
\mathbb P^{\mbox{i}\pm}(N,\vec v)\equiv \left\{{\cal P}\in\mathscr P(N):\;\mathbb F^{\mbox{i}\pm}(N,\vec v,{\cal P})\neq\emptyset\right\}.
\end{equation}
Yang \cite{Y22} demonstrated that 
\begin{equation}\label{qi-star}
\mathbb Q^{\mbox{i}\pm}(N,\vec v)=\bigcup_{{\cal P}\in \mathbb P^{\mbox{i}\pm}(N,\vec v)}\{{\cal P}\}\times\mathbb F^{\mbox{i}\pm}(N,\vec v,{\cal P}).
\end{equation}
Moreover, the ${\cal P}$-set defined for~(\ref{qu}) also satisfies
\begin{equation}\label{pu-def}
\mathbb P^{\mbox{u}}(N,\vec v)=\left\{{\cal P}\in\mathscr P(N):\;\tilde w(N,\vec v,{\cal P})\geq \tilde w(N,\vec v,{\cal P}'),\;\forall {\cal P}'\in\mathscr U(N,{\cal P})\right\}.
\end{equation}

A combination of~(\ref{stab-def}),~(\ref{qu}), and~(\ref{qi-star}) would lead to
\begin{equation}\label{stab-def-niu}
\mathbb S^\pm(N,\vec v)=\bigcup_{{\cal P}\in \mathbb P^{\mbox{i}\pm}(N,\vec v)\cap\mathbb P^{\mbox u}(N,\vec v)}\{{\cal P}\}\times \mathbb F^{\mbox{i}\pm}(N,\vec v,{\cal P}).
\end{equation}
From~(\ref{core-relations}) and~(\ref{patch-up}), we can derive that
\begin{equation}\label{patchup-relations}
\mathbb F^{\mbox{i}+}(N,\vec v,{\cal P})
\subseteq \mathbb F^{\mbox{i}-}(N,\vec v,{\cal P}).
\end{equation}
Plugging~(\ref{patchup-relations}) into~(\ref{pistar-def}), it is easy to see that
\begin{equation}\label{pi-relations}
\mathbb P^{\mbox{i}+}(N,\vec v)\subseteq 
\mathbb P^{\mbox{i}-}(N,\vec v).
\end{equation}
With~(\ref{stab-def-niu}) to~(\ref{pi-relations}), we would certainly have the following relationships:
\begin{equation}\label{stab-relations}
\mathbb S^+(N,\vec v)\subseteq 
\mathbb S^-(N,\vec v).
\end{equation}
 
As an intermediate concept $\mathbb S^0$ is already universal,~(\ref{stab-relations}) would lead to
\begin{equation}\label{univ-medium}
\mathbb S^-(N,\vec v)\neq\emptyset,\hspace*{.8in}\forall \vec v\in \mathscr V_{++}(N).
\end{equation}
On the other hand, the core-based strong stability $\mathbb S^+$ is easily shown to be not universal. There are also the following compatibilities: 
\begin{equation}\label{badaling}\begin{array}{l}
\mathbb S^\pm(N,\vec v)\cap \left[\{\{N\}\}\times \partial(N,\vec v)\right]=\mathbb Q^{\mbox{i}\pm}(N,\vec v)\cap \left[\{\{N\}\}\times \partial(N,\vec v)\right]\\
\;\;\;\;\;\;\;\;\;\;\;\;\;\;\;\;\;\;\;\;\;\;\;\;\;\;\;\;\;\;\;\;\;\;\;\;\;\;\;\;\;\;\;\;\;\;\;=\{\{N\}\}\times \mathbb F^\pm(N,\vec v).
\end{array}\end{equation}

We now revisit the comparative statics of Section~\ref{ordering} concerning changing $\vec v$'s. Suppose $\vec v^{\,1}\leq_{\mbox{cp}}\vec v^{\,2}$ as defined by~(\ref{partial-order}). Then due to~(\ref{pistar-def}) and~(\ref{qi-star}), we could tell from Theorem~\ref{t-relation} that 
\begin{equation}\label{java}
\mathbb P^{\mbox{i}\pm}(N,\vec v^{\,1})\subseteq  \mathbb P^{\mbox{i}\pm}(N,\vec v^{\,2});
\end{equation}
also, for any ${\cal P}\in \mathbb P^{\mbox{i}\pm}(N,\vec v^{\,1})\subseteq \mathbb P^{\mbox{i}\pm}(N,\vec v^{\,2})$,
\begin{equation}\label{sumatra}
\mathbb F^{\mbox{i}\pm}(N,\vec v^{\,1},{\cal P})\subseteq \mathbb F^{\mbox{i}\pm}(N,\vec v^{\,2},{\cal P}).
\end{equation}
Since $\mathscr I(N,{\cal P}')\subset \mathscr I(N,{\cal P})$ whenever ${\cal P}'$ is itself in ${\cal P}$'s fission-down-to neighborhood,~(\ref{fi-resist-weak}) to~(\ref{fi-resist-strong}) would make it ``easier'' for a ``more splintered'' partition to be fission-resistant; also,~(\ref{fu-resist}) would make it ``easier'' for a ``more consolidated'' partition to be fusion-resistant. 

\begin{figure}[h]
\centering
\includegraphics[width=1.05\columnwidth]{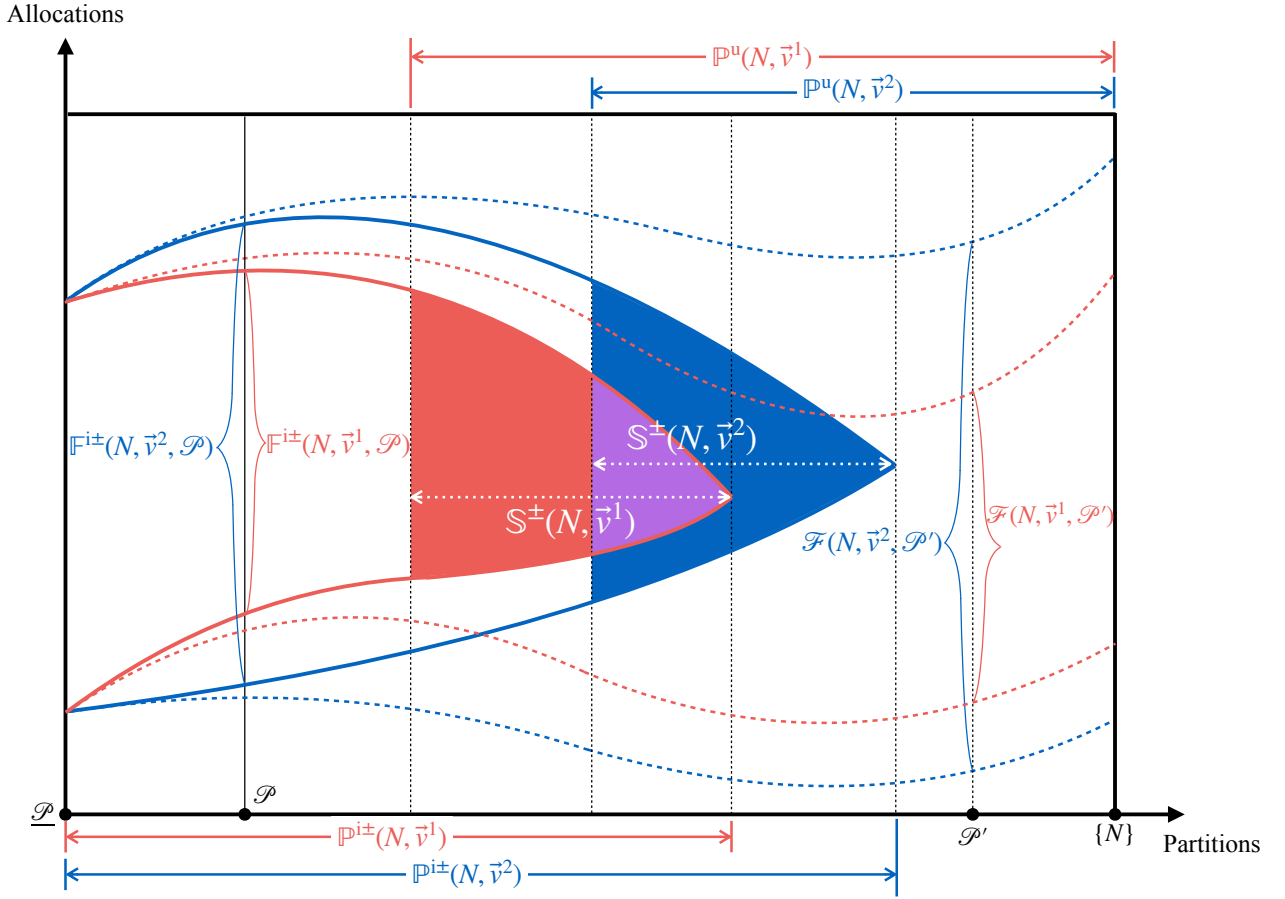}
\caption{A Schematic Sketch of the Various Entities under Two Ordered Games}\label{fig2}
\end{figure}

When ${\cal P}$ happens to be the all-consolidated partition $\{N\}$,~(\ref{sumatra}) would convey the ``more cooperation'' message of $\mathbb F^\pm(N,\vec v^{\,1}) \subseteq \mathbb F^\pm(N,\vec v^{\,2})$. The $+$ case would be Yang and Li's \cite{YL20} core-centric theme of $\mathbb F^+(N,\vec v^{\,1}) \subseteq \mathbb F^+(N,\vec v^{\,2})$. With~(\ref{java}), our current message is more general. It also involves the shifting to the ``more consolidated'' end of partitions. Combine these rough understandings with Theorem~\ref{t-relation}, and we can come to Figure~\ref{fig2}'s depiction of the various entities under two games that are ordered centripetally. 

We are conscious that either $\mathscr F(N,\vec v^i,\underline{\cal P})$ for the all-splintered partition $\underline{\cal P}$ has exactly one member $\vec 1$; also, $\mathscr F(N,\vec v^i,{\cal P})$ could be empty for some partitions ${\cal P}$. Moreover, the partitions are by no means totally ordered, nor are the sets $\mathbb P^{\mbox{i}\pm}(N,\vec v^i)$ and $\mathbb P^{\mbox{u}}(N,\vec v^i)$ necessarily connected. We have also drawn a case where $\mathbb S^\pm(N,\vec v^i)$ is nonempty. This might not be the case for the strong concept $\mathbb S^+$. Nevertheless, 
Figure~\ref{fig2} has reflected Theorem~\ref{t-relation} along with~(\ref{java}) and~(\ref{sumatra}). From the figure, we may discern that stable solutions tend to be ``more consolidated'' when the underlying game becomes more centripetal. 

On the one extreme end, $\vec f\in \mathbb F^\pm(N,\vec v^{\,1})$ would happen when $\mathbb Q^{\mbox{i}\pm}(N,\vec v^{\,1})$ extends to the ``rightmost'' corner; the latter would certainly be covered by the more expansive $\mathbb Q^{\mbox{i}\pm}(N,\vec v^{\,2})$ while also automatically covered by any $\mathbb Q^{\mbox u}(N,\vec v^{\,2})$. Hence, $\vec f\in \mathbb F^*(N,\vec v^{\,2})$ as well. On the symmetric but much less visited extreme end, we may also tell from the figure that the pair made up of the all-splintered partition $\underline{\cal P}$ and the all-one vector $\vec 1$ which is the only member of $\mathscr F(N,\vec v,\underline{\cal P})$ would be stable for $(N,\vec v^{\,1})$ when it is so for $(N,\vec v^{\,2})$. More rigorously, both observations can be verified by tapping into Theorem~\ref{t-relation}. 

\begin{corollary}\label{c-extremes}
Suppose $\vec v^{\,1},\vec v^{\,2}\in\mathscr V_{++}(N)$ satisfy $\vec v^{\,1}\leq_{\mbox{cp}} \vec v^{\,2}$ in the sense of~(\ref{partial-order}). Then, 
\[ \mathbb F^\pm(N,\vec v^{\,1})\subseteq \mathbb F^\pm(N,\vec v^{\,2});\]
\[ (\underline{\cal P},\vec 1)\in \mathbb S^\pm(N,\vec v^{\,2})\;\;\;\Longrightarrow\;\;\;  (\underline{\cal P},\vec 1)\in \mathbb S^\pm(N,\vec v^{\,1}). \]
\end{corollary} 	

\noindent{\bf Proof of Corollary~\ref{c-extremes}: }Suppose $\vec f\in \mathbb F^\pm(N,\vec v^{\,1})$. By~(\ref{badaling}), this amounts to 
\begin{equation}\label{pos2}
\mathbb Q^{\mbox{i}\pm}(N,\vec v^{\,1})\cap \left[\{\{N\}\}\times \partial(N,\vec v^{\,1})\right]=\{\{N\}\}\times \mathbb F^\pm(N,\vec v^{\,1})\ni (\{N\},\vec f).
\end{equation}
Now Theorem~\ref{t-relation} states that $\partial(N,\vec v^{\,1})\subseteq \partial(N,\vec v^{\,2})$ and $\mathbb Q^{\mbox{i}\pm}(N,\vec v^{\,1})\subseteq \mathbb Q^{\mbox{i}\pm}(N,\vec v^{\,2})$. Thus, the above~(\ref{pos2}) would be equivalent to 
\begin{equation}\label{pos22}
\{\{N\}\}\times \mathbb F^\pm(N,\vec v^{\,2})=\mathbb Q^{\mbox{i}\pm}(N,\vec v^{\,2})\cap \left[\{\{N\}\}\times \partial(N,\vec v^{\,2})\right]\ni (\{N\},\vec f).
\end{equation}
Hence, $f\in \mathbb F^\pm(N,\vec v^{\,2})$ as well. 

Next, suppose $(\underline{\cal P},\vec 1)\in \mathbb S^\pm(N,\vec v^{\,2})$. Due to~(\ref{fi-resist-weak}) to~(\ref{fi-resist-strong}), it is already true that 
\begin{equation}\label{useful}
(\underline{\cal P},\vec 1)\in \mathbb Q^{\mbox{i}\pm}(N,\vec v^{\,1})\cap \mathbb Q^{\mbox{i}\pm}(N,\vec v^{\,2})
\end{equation}
In view of~(\ref{stab-def}),~(\ref{qu}), and~(\ref{useful}),
\begin{equation}\label{sop1}
\underline{\cal P}\in \mathbb P^{\mbox u}(N,\vec v^{\,2}).
\end{equation}
Since Theorem~\ref{t-relation} states that $\mathbb P^{\mbox u}(N,\vec v^{\,2})\subseteq \mathbb P^{\mbox u}(N,\vec v^{\,1})$,~(\ref{sop1}) would imply 
\begin{equation}\label{sop2}
\underline{\cal P}\in \mathbb P^{\mbox u}(N,\vec v^{\,1}).
\end{equation}
By~(\ref{stab-def}),~(\ref{qu}), and~(\ref{useful}), the above~(\ref{sop2}) would amount to $(\underline{\cal P},\vec 1)\in \mathbb S^\pm(N,\vec v^{\,1})$. \qed

With Theorem~\ref{t-relation} and Corollary~\ref{c-extremes}, we have thus completed the second link in~(\ref{logic}), from ``centripetality'' to ``consolidation/cooperation''.  

\section{From Risk Aversion and Onward}\label{inoutreward}



For applications of~(\ref{logic})'s second link, let there be a probability space $(\Omega,\mathscr G,p)$. Denote $\int_\Omega Y(\omega)\cdot p(d\omega)$ for any random variable $Y$ defined on the space by $\mathbb E[Y]$. Also, let $\mathbb L^2$ be the space of random variables $Y$ with $\mathbb E[|Y|^2]<+\infty$. Suppose each coalition $C$ is associated with a random outcome $\Phi(C)\in \mathbb L^2$. Also, suppose all players share a common reward function $\tilde R:\mathbb L^2\rightarrow \Re_{++}$ that is positively homogeneous in the sense that
\begin{equation}\label{homo}
\tilde R(f\cdot Y)=f\cdot \tilde R(Y),\hspace*{.8in}\forall f\geq 0,\;Y\in \mathbb L^2.
\end{equation}

For a coalition $C$, each $\vec f\equiv (f(i))_{i\in C}$ constitutes a plan to divide the coalition's outcome that has been agreed upon prior to the outcome's actual fruition. Note player $i$'s gain in a coalition $C\ni i$ under any share $f(i)$ would be $\tilde R\left(f(i)\cdot \Phi(C)\right)$ which due to~(\ref{homo}) is merely $f(i)\cdot \tilde R(\Phi(C))$. Then, our situation 
can now be understood as a coalitional game $(N,\vec v)$ with 
$\vec v\equiv (v(C))_{C\in\mathscr C(N)}$ satisfying
\begin{equation}\label{value-def}
v(C)\equiv \tilde R(\Phi(C)),\hspace*{.8in}\forall C\in\mathscr C(N).
\end{equation}
For two special occasions, 
we can apply Theorem~\ref{t-relation} and Corollary~\ref{c-extremes} to the game $(N,\vec v)$. 


In the first situation, we suppose players $i\in N$ have with them independent and identically distributed inputs $\Theta(i)\in \mathbb L^2$ with a common mean $\bar\mu\equiv \mathbb E[\Theta(i)]>0$ and common standard deviation $\bar\sigma\equiv \sqrt{\mathbb E[\Theta(i)^{\,2}]-(\mathbb E[\Theta(i)])^2}>0$. Each coalition $C\in \mathscr C(N)$ is also associated with some $\phi(C)\in \Re_{++}$ so that 
\begin{equation}\label{outcome-special}
\Phi(C)\equiv \phi(C)\cdot 
\sum_{i\in C}\Theta(i). 
\end{equation}
When the $\phi(\{i\})$'s all equal to one, we may understand each $\phi(C)$ in~(\ref{outcome-special}) as a certain amplifier on the total effect of all participating players' individual contributions. When $\phi(C)>1$ which could typically happen when $C$ is not too large, it might be said that by banding together into $C$, the participating players can achieve more than their sum due to some synergistic effects; when $\phi(C)<1$ which could happen when $C$ is too large, internecine frictions or organizational complexities could finally catch up with any economy of scale. 

At some $\bar r\in [0,\bar\mu/\bar\sigma)$, we suppose the reward $\tilde R$ satisfies
\begin{equation}\label{reward-special}
\tilde R(Y)\equiv\mathbb E[Y]-\bar r\cdot\sqrt{\mathbb E[Y^2]-(\mathbb E[Y])^2},\hspace*{.8in}\forall Y\in \mathbb L^2.
\end{equation} 
For any parameters $\bar r^k$ for $k=1,2$, let us use $\tilde R^k$ for the reward function resulting from~(\ref{reward-special}) and then $\vec v^{\,k}$ for the value vector resulting from~(\ref{value-def}) and~(\ref{outcome-special}).  

\begin{proposition}\label{p-pity}
For the games $(N,\vec v^{\,k})$ defined through~(\ref{value-def}) to~(\ref{reward-special}) for $k=1,2$, we would have~(\ref{partial-order}) when $\bar r^1\leq \bar r^2$. 
\end{proposition}

\noindent{\bf Proof of Proposition~\ref{p-pity}: }Due to the description from~(\ref{value-def}) to~(\ref{reward-special}), 
\begin{equation}\label{bagua}
v^k(C)=|C|\cdot\bar\mu-\bar r^k\cdot \sqrt{|C|}\cdot\bar\sigma,\hspace*{.8in}\forall C\in \mathscr C(N),\;k=1,2.
\end{equation}
For $C^1,C^2\in \mathscr C(N)$ with $C^1\subseteq C^2$, we have $|C^2|\geq|C^1|\geq 1$. But when combined with the fact that $0\leq \bar r^1\leq\bar r^2<\bar\mu/\bar\sigma$, this would lead to 
\begin{equation}\label{gabaj}
\frac{|C^2|\cdot\bar\mu-\bar r^1\cdot \sqrt{|C^2|}\cdot\bar\sigma}{|C^1|\cdot\bar\mu-\bar r^1\cdot \sqrt{|C^1|}\cdot\bar\sigma}\leq \frac{|C^2|\cdot\bar\mu-\bar r^2\cdot \sqrt{|C^2|}\cdot\bar\sigma}{|C^1|\cdot\bar\mu-\bar r^2\cdot \sqrt{|C^1|}\cdot\bar\sigma}.
\end{equation}
In view of~(\ref{bagua}), the above~(\ref{gabaj}) is just~(\ref{partial-order}). \qed

When Theorem~\ref{t-relation}, Corollary~\ref{c-extremes}, and Proposition~\ref{p-pity} are put together, we would certainly get the message that ``aversion to risk promotes cooperation''. 

Our reward function of~(\ref{reward-special}) is indeed law-invariant. Rather than the $\Phi(C)$'s, we might as well use the $\tilde k(C)$'s to characterize the coalitions, where each $\tilde k(C)$ is the inverse of the corresponding $\Phi(C)$'s cumulative distribution function: 
\[ \tilde k(C,\alpha)\equiv \inf\{\phi\in\Re:\;p(\Phi(C)\leq \phi)\geq \alpha\},\hspace*{.8in}\forall 
\alpha\in [0,1].\]
Let $\mathscr K$ be the space of all such quantile functions. We can think of the reward function as some $\tilde r:\mathscr K\rightarrow \Re_{++}$ instead of $\tilde R:\mathbb L^2\rightarrow \Re_{++}$. Then, instead of~(\ref{value-def}), we might adopt
\begin{equation}\label{value-redef}
v(C)\equiv \tilde r(\tilde k(C)),\hspace*{.8in}\forall C\in \mathscr C(N).
\end{equation}
Though not that helpful to the aforementioned occasion, we find~(\ref{value-redef}) to be more useful for another situation involving different outcome and reward formations. 

\section{Another Law-invariant Situation}\label{situ2}

\begin{figure}[h]
\centering
\includegraphics[width=0.85\columnwidth]{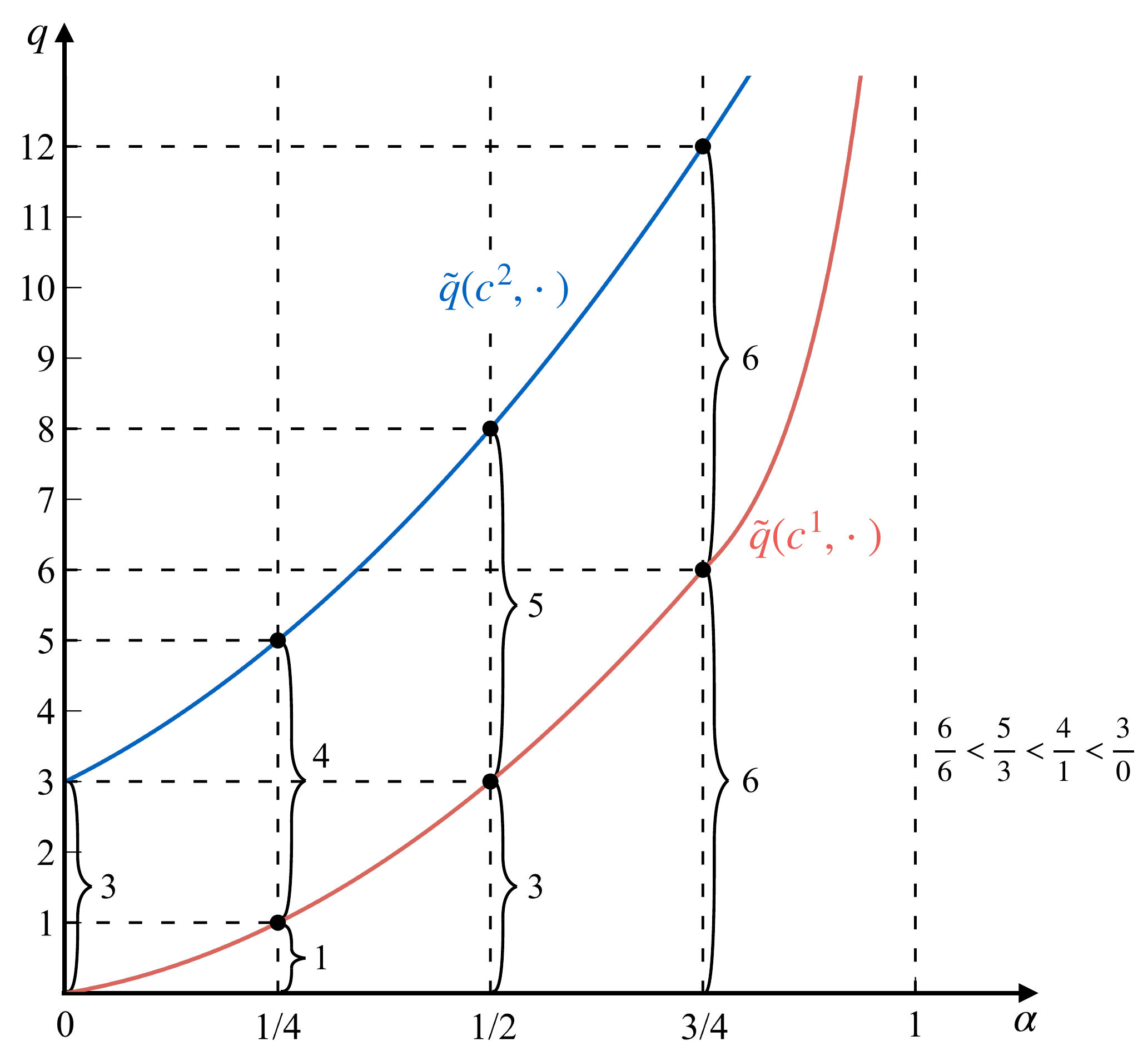}
\caption{An Illustrative Example of Quantile Functions}\label{f-123456}
\end{figure}

Here, we suppose each $\tilde k(C,\cdot)$ is continuous and each $\tilde k(C,\alpha)$ is strictly positive as long as $\alpha>0$. For the entire outcome-quantile vector $(\tilde k(C))_{C\in\mathscr C(N)}\in \mathscr K^{\mathscr C(N)}$, we suppose 
\begin{equation}\label{cond}
\frac{\tilde k(C^2,\alpha^2)}{\tilde k(C^1,\alpha^2)}\leq \frac{\tilde k(C^2,\alpha^1)}{\tilde k(C^1,\alpha^1)},\hspace*{.5in}\forall C^1,C^2\in \mathscr C(N)\mbox{ with }C^1\subseteq C^2\mbox{ and }0<\alpha^1\leq\alpha^2<1. 
\end{equation}
According to~(\ref{cond}), a larger coalition $C^2$, when compared to a smaller coalition $C^1$, would have a relatively higher $\alpha$-quantile as $\alpha$ becomes smaller say going from 10\% to 1\%; see Figure~\ref{f-123456} for an illustrative example. 

It is clear from Figure~\ref{f-123456} that~(\ref{cond}) is one way of stating that a larger coalition contains relatively less downside risk. For instance, when $\Phi(C)$ is uniformly distributed between some $\underline \phi(|C|)>0$ and some $\overline \phi(|C|)$ which is above $\underline\phi(|C|)$, we would have
\[ \tilde k(C,\alpha)=(1-\alpha)\cdot\underline \phi(|C|)+\alpha\cdot\overline \phi(|C|).\]
Then,~(\ref{cond}) would amount to the decrease of $\overline \phi(s)/\underline \phi(s)$ in the coalition size $s=1,2,...$. This just means that as the coalition $C$ grows, its random outcome $\Phi(C)$ would be ``less spread-out'' in a relative sense that is  characterizable by a shrinking $\overline\phi(|C|)/\underline \phi(|C|)$. Much like the earlier independent-input case but expressed differently, this somehow reflects a larger coalition's increased ability to have extremes of its participants canceled out with each other.    

For the reward function $\tilde r$ to be used at~(\ref{value-redef}), we let it be parameterized by a continuous and strictly positive probability density function $\bar\mu$ on $[0,1]$ so that  
\begin{equation}\label{mixed-avr}
\tilde r(k)\equiv \int_0^1\bar a(k,\alpha)\cdot \bar\mu(\alpha)\cdot d\alpha,\hspace*{.8in}\forall k\in\mathscr K,
\end{equation}
where $\bar a(\cdot,\alpha)$ is the $\alpha$-level conditional value at risk defined as in
\begin{equation}\label{avr}
\bar a(k,\alpha)\equiv \frac{1}{1-\alpha}\cdot \int_0^{1-\alpha}k(\beta)\cdot d\beta,\hspace*{.8in}\forall k\in\mathscr K,\;\alpha\in [0,1).
\end{equation}
If instead of the current one concerned with one single $\bar\mu$, the function $\tilde r(\cdot)$ becomes the minimum of~(\ref{mixed-avr})'s right-hand side over a set ${\cal M}$ of $\bar\mu$'s, then we would be dealing with a general law-invariant coherent risk measure; see Kusuoka \cite{K01}.

Our~(\ref{mixed-avr}) and~(\ref{avr}) have not yet reached such generality. Still, they are adequate to characterize risk aversion. When $\alpha$ increases in~(\ref{avr}), the reward $\bar a(\cdot,\alpha)$ would focus on the narrower band $[0,1-\alpha]$ corresponding to to the more narrowly defined worst $(1-\alpha)$-portion of the random outcome as reflected in $k$; when $\bar\mu$ shifts more towards larger $\alpha$ values in~(\ref{mixed-avr}), the resulting $\tilde r(\cdot)$ would still care more about the lower end of the random outcome. 

For our ensuing derivations which rely heavily on Karlin and Rinott's \cite{KR80} preservation result related somewhat to log-supermodularity, we find it convenient to express $\bar\mu$'s shifts in the likelihood ratio sense. We consider $\bar\mu^1\leq_{\mbox{lr}}\bar\mu^2$ when
\begin{equation}\label{lr-ranking}
\frac{\bar\mu^1(\alpha^2)}{\bar\mu^1(\alpha^1)}\leq \frac{\bar\mu^2(\alpha^2)}{\bar\mu^2(\alpha^1)},\hspace*{.8in}\forall 0<\alpha^1\leq\alpha^2<1.
\end{equation}  
For any parameters $\bar\mu^k$ for $k=1,2$, let us use $\tilde r^k$ for the reward function resulting from~(\ref{mixed-avr}) and~(\ref{avr}) and then $\vec v^{\,k}$ for the value vector resulting from~(\ref{value-redef}) and~(\ref{cond}).  

\begin{proposition}\label{p-okla}
For the games $(N,\vec v^{\,k})$ defined through~(\ref{value-redef}) to~(\ref{avr}) for $k=1,2$, we would have~(\ref{partial-order}) when $\bar\mu^1\leq_{\mbox{lr}} \bar \mu^2$ in the sense of~(\ref{lr-ranking}). 
\end{proposition}

\noindent{\bf Proof of Proposition~\ref{p-okla}: }Fix some $C^1,C^2\in \mathscr C(N)$ with $C^1\subseteq C^2$. We first show the intermediate result that under~(\ref{cond}) and~(\ref{avr}), 
\begin{equation}\label{one-unit}
\frac{\bar a(\tilde k(C^2),\alpha^1)}{\bar a(\tilde k(C^1),\alpha^1)}\leq \frac{\bar a(\tilde k(C^2),\alpha^2)}{\bar a(\tilde k(C^1),\alpha^2)},\hspace*{.8in}\forall 0<\alpha^1\leq\alpha^2<1.
\end{equation}

To this end, define  $z_1(\cdot)=z_3(\cdot)=\tilde k(C^1,\cdot)$ and $z_2(\cdot)=z_4(\cdot)=\tilde k(C^2,\cdot)$. Due to~(\ref{cond}), 
\begin{equation}\label{use001}
z_1(\beta)\cdot z_2(\beta')\leq z_3(\beta\vee \beta')\cdot z_4(\beta\wedge \beta').
\end{equation}
Now for any $0<\alpha^1\leq\alpha^2<1$, define $g_1(\cdot)=g_4(\cdot)={\bf 1}_{(0,1-\alpha^2]}(\cdot)$ and $g_2(\cdot)=g_3(\cdot)={\bf 1}_{(0,1-\alpha^1]}(\cdot)$. It is easy to see that 
\begin{equation}\label{use002}
g_1(\beta)\cdot g_2(\beta')\leq g_3(\beta\vee \beta')\cdot g_4(\beta\wedge \beta').
\end{equation}

Combine~(\ref{use001}) and~(\ref{use002}), and we arrive to
\begin{equation}\label{used-0}
(z_1g_1)(\beta)\cdot (z_2g_2)(\beta')\leq (z_3g_3)(\beta\vee \beta')\cdot (z_4g_4)(\beta\wedge \beta').
\end{equation}
By Theorem 2.1 of Karlin and Rinott \cite{KR80},~(\ref{used-0}) would result in 
\begin{equation}\label{use-0}
\left[\int_{0}^1(z_1g_1)(\beta)\cdot d\beta\right]\cdot\left[\int_0^1 (z_2g_2)(\beta)\cdot d\beta\right]\leq \left[\int_0^1(z_3g_3)(\beta)\cdot d\beta\right]\cdot\left[\int_0^1 (z_4g_4)(\beta)\cdot d\beta\right].
\end{equation}
But by the definitions of the $f(i)$'s and $g(i)$'s,~(\ref{use-0}) is just
\begin{equation}\label{menkan}
\frac{\int_0^{1-\alpha^1}\tilde k(C^2,\beta)\cdot d\beta}{\int_0^{1-\alpha^1}\tilde k(C^1,\beta)\cdot d\beta}\leq \frac{\int_0^{1-\alpha^2}\tilde k(C^2,\beta)\cdot d\beta}{\int_0^{1-\alpha^2}\tilde k(C^1,\beta)\cdot d\beta}.
\end{equation}

In view of~(\ref{avr}), the above~(\ref{menkan}) would just mean~(\ref{one-unit}). Now let $z_1(\alpha)=\bar a(\tilde k(C^1),\alpha)\cdot \bar\mu^2(\alpha)$, $z_2(\alpha)=\bar a(\tilde k(C^2),\alpha)\cdot \bar\mu^1(\alpha)$, $z_3(\alpha)=\bar a(\tilde k(C^2),\alpha)\cdot \bar\mu^2(\alpha)$, and $z_4(\alpha)=\bar a(\tilde k(C^1),\alpha)\cdot \bar\mu^1(\alpha)$. 

Then,~(\ref{lr-ranking}) and~(\ref{one-unit}) would together imply
\begin{equation}\label{pre-karlin}
z_1(\alpha)\cdot z_2(\alpha')\leq z_3(\alpha\vee \alpha')\cdot z_4(\alpha\wedge \alpha').
\end{equation}
By Theorem 2.1 of Karlin and Rinott \cite{KR80},~(\ref{pre-karlin}) would lead to
\begin{equation}\label{dada}
\left[\int_0^1 z_1(\alpha)\cdot d\alpha\right]\cdot \left[\int_0^1 z_2(\alpha)\cdot d\alpha\right]\leq \left[\int_0^1 z_3(\alpha)\cdot d\alpha\right]\cdot \left[\int_0^1 z_4(\alpha)\cdot d\alpha\right].
\end{equation}
By the definitions of the $f(i)$'s,~(\ref{dada}) just means that
\begin{equation}\label{lala}
\frac{\int_0^1\bar a(\tilde k(C^2),\alpha)\cdot \mu^1(\alpha)\cdot d\alpha}{\int_0^1\bar a(\tilde k(C^1),\alpha)\cdot \mu^1(\alpha)\cdot d\alpha}\leq \frac{\int_0^1\bar a(\tilde k(C^2),\alpha)\cdot \mu^2(\alpha)\cdot d\alpha}{\int_0^1\bar a(\tilde k(C^1),\alpha)\cdot \mu^2(\alpha)\cdot d\alpha}.
\end{equation}
In view of~(\ref{mixed-avr}), the above~(\ref{lala}) would just be
\begin{equation}\label{navy}
\frac{\tilde r^1(\tilde k(C^2))}{\tilde r^1(\tilde k(C^1))}\leq \frac{\tilde r^2(\tilde k(C^2))}{\tilde r^2(\tilde k(C^1))}.
\end{equation}
By~(\ref{value-redef}) further,~(\ref{navy}) would amount to~(\ref{partial-order}). \qed

Both Propositions~\ref{p-pity} and~\ref{p-okla} are fulfillments of the first link in~(\ref{logic}), from ``risk aversion'' to ``centripetality''. Proposition~\ref{p-okla} provides another channel for Theorem~\ref{t-relation} and Corollary~\ref{c-extremes} to deliver the message that ``aversion to risk promotes cooperation''. 

\section{Concluding Remarks}\label{conclusion}

The first link in~(\ref{logic}) awaits further strengthenings. So far, two quite disjoint cases have been treated. Their generalizations or even better, unification, would be more than welcome. Besides, the fractional setup is suitable for pre-determined allocations. More complex situations might arise when players could re-negotiate their shares after the revelations of the coalitional outcomes.   
The second link in~(\ref{logic}) from centripetality to cooperation 
might have other applications beyond anything to do with uncertainty or risk. Further improvements to the link itself would surely be desirable. For instance, it would be good to have the restriction on strictly positive values lifted.  





\end{document}